\documentclass{aa}
\usepackage{natbib}
\bibpunct{(}{)}{;}{a}{}{,}
\usepackage{graphicx}
\graphicspath{{Images/}}
\usepackage{txfonts}
\usepackage{subfig}
\usepackage{array}
\usepackage{hyperref}
\usepackage{makecell}
\usepackage{lastpage}
\hypersetup{
    colorlinks=true,
    linkcolor=blue,     
    urlcolor=blue,
    citecolor=blue
    }

\def\NH{$N$\textsubscript{H}}
\def\NHeq{$N$\textsubscript{H,eq}}
\def\ER{$\lambda_{\rm{Edd}}$}
\newcommand{\PaperI}{\hyperlink{cite.refId0}{Paper~I}}

\begin{document} 

\title{Population synthesis of active galactic nuclei based on the radiation-regulated unification model}

\author{Dimitra Gerolymatou\inst{1}\corrauth{dimitra.gerolymatou@unige.ch}, Stéphane Paltani\inst{1}\email{stephane.paltani@unige.ch}, Claudio Ricci\inst{1}\email{claudio.ricci@unige.ch}, Tonima T. Ananna\inst{2}\email{tonimatasnim@gmail.com}}

\institute{Department of Astronomy, University of Geneva, 1290 Versoix, Switzerland \and Department of Physics and Astronomy, Wayne State University, USA}

 
\abstract
   {X-ray surveys of active galactic nuclei (AGNs) provide direct constraints on the properties of individual AGNs, such as their emission, obscuration, and accretion rate. Previous AGN population synthesis models have not addressed such properties self-consistently.
   Here, we use a simulation-based inference (SBI) approach to constrain the geometrical and physical properties of the AGN population. We perform numerical simulations with our ray-tracing code, \textsc{RefleX}, which allows the self-consistent modelling of the X-ray emission of AGNs with flexible circumnuclear and source geometries.
   We create our synthetic population by sampling the intrinsic active black hole mass function (BHMF) and Eddington ratio distribution function (ERDF) of local AGNs, and we construct a geometry based on the radiation-regulated model, along with Eddington-ratio-dependent emission spectra. Using the \textsc{RefleX}-simulated emission of the AGN population, we aim to simultaneously reproduce the cosmic X-ray background (CXB), differential AGN number counts, and several observed absorption properties of local AGNs, such as the fraction of \NH{} in bins of log(\NH{}), the Compton-thick fraction as a function of limiting flux, and the number of obscured and unobscured AGNs as a function of Eddington ratio. With this approach, we test the consistency of the radiation-regulated model with a very comprehensive set of X-ray observables, while constraining the size and density of the dusty torus and the evolution of the local AGN population.
  We derive an intrinsic Compton-thick fraction of $40\pm3$\%, and find that a simple evolutionary prescription controlling the active fraction of supermassive black holes is sufficient for our synthetic population to reproduce the CXB. Finally, we compare the inferred dusty torus structure to observations and examine the dependence of the covering factor on Eddington ratio, the observed correlation between reflection and obscuration, and the BHMFs and ERDFs of obscured and unobscured AGNs.
  }

\keywords{galaxies: active -- X-rays: diffuse background}

\titlerunning{AGN population synthesis based on radiation-regulated unification}
\authorrunning{D. Gerolymatou et al}
\maketitle
\nolinenumbers

\section{Introduction} \label{sec:intro}
The supermassive black holes (SMBHs) at the galaxy centres, powering active galactic nuclei (AGNs), have been found to be linked to key properties of their host galaxies such as mass, velocity dispersion, and luminosity \citep[e.g.,][]{1998AJ....115.2285M, 2000ApJ...539L..13G, 2013ARA&A..51..511K}. The relationships between SMBH masses and these properties suggest that SMBHs may play an important role in galaxy evolution. To investigate how SMBHs evolve and interact with their host galaxies, extensive population studies of AGNs have been conducted, particularly using X-ray observations. These studies allow us to derive the X-ray luminosity function \citep[XLF; e.g., more recently in][]{2014ApJ...786..104U, 2015MNRAS.451.1892A, 2019ApJ...871..240A}, the active black hole mass function (BHMF), and the Eddington ratio distribution function (ERDF) \citep[e.g.][]{2013MNRAS.428..421S,2015MNRAS.447.2085S,2018MNRAS.474.1225A, 2022ApJS..261....9A, 2024ApJ...962..152H}.

X-ray emission is a fundamental signature of AGNs, consistently outshining the emission from the host galaxies themselves \citep[e.g.,][]{2015A&ARv..23....1B}. X-ray observations are crucial for detecting obscured AGNs, as X-ray photons can penetrate the dense dusty structures surrounding the SMBHs and provide us with valuable information about the innermost regions of AGNs. Primary X-ray emission is attributed to the inverse Compton scattering of accretion disc photons within a hot corona close to the SMBH \citep[e.g.,][]{1991ApJ...380L..51H}. Reprocessed emission is due to reflection on the circumnuclear material \citep[e.g.,][]{1990Natur.344..132P, 1994MNRAS.267..743G}. Absorption by the circumnuclear material substantially shapes the observed X-ray emission spectrum and thus the modelling of the AGN population. The parsec-scale structure around AGNs, commonly referred to as the dusty torus \citep[e.g.,][]{2015ARA&A..53..365N,2018ARA&A..56..625H}, acts simultaneously as an absorber and reflector. Other structures such as the broad-line region (BLR) can also contribute to absorption and reflection, while the accretion disc acts as a reflector.

X-ray surveys provide important insights into the obscuring material around the central engine of AGNs. One key result obtained by AGN surveys over the past decades is that the fraction of obscured AGNs (typically defined as the fraction of AGN with \NH{} $\geq 10^{22}$ cm$^{-2}$ within the entire population) is anti-correlated with luminosity \citep[e.g.,][]{2003ApJ...598..886U, 2011ApJ...728...58B, 2015ApJ...815L..13R}. This implies that the covering factor of the circumnuclear material depends on luminosity, which cannot be explained by the simple unification of AGN \citep{1993ARA&A..31..473A}, according to which the observational differences between AGNs arise solely due to orientation. The receding torus model was proposed to explain the luminosity dependence \citep{1991MNRAS.252..586L}, but more recent studies propose a radiation-regulated model instead \citep{2017Natur.549..488R, 2022ApJ...938...67R}. In the radiation-regulated unification model, the driving parameter of the covering factor is the Eddington ratio rather than luminosity, with radiation pressure from the accreting SMBH regulating obscuration levels by clearing the circumnuclear environment.

The intrinsic fraction of Compton-thick AGNs (CTK; typically defined as the fraction of AGNs with \NH{} $\geq 10^{24}$ cm$^{-2}$ within the entire population) remains a subject of debate, with reported values from population synthesis models ranging from 10 to 50\% \citep[e.g.,][]{2009ApJ...696..110T, 2012A&A...546A..98A, 2014ApJ...786..104U, 2015MNRAS.451.1892A, 2019ApJ...871..240A}. Direct measurements by hard X-ray surveys in the local Universe find an intrinsic CTK fraction from 20 to 35\% \citep[e.g.,][]{2015ApJ...815L..13R, 2017ApJ...846...20L,2025ApJ...978..118B}.

These previous population synthesis studies have treated reflection and absorption in AGNs as independent factors, adopting templates for the observed X-ray spectra. However, to accurately model AGNs, a self-consistent approach that connects emission, obscuration, and reflection with geometry is essential. In \cite{refId0} (hereafter Paper I), we developed an AGN population synthesis framework in which obscuration and reflection are intrinsically linked through the spatial distribution of the circumnuclear material. Employing a luminosity-dependent torus we reproduced the cosmic X-ray background (CXB) and observed AGN absorption properties simultaneously.

In this work, we extend the methodology of \PaperI{} by replacing the XLF with the ERDF and BHMF and constructing a model based on the radiation-regulated unification of AGNs. This way, we advance population synthesis from phenomenological to physical modelling. Specifically, using numerical simulations with the \textsc{RefleX} ray-tracing code \citep{2017A&A...607A..31P}, we simulate the X-ray emission of AGNs with various geometries and emission spectra. We generate a synthetic AGN population by sampling the local intrinsic active BHMF and corresponding ERDF coupled with a simple prescription for their evolution, and we build a geometry informed by the radiation-regulated model. With a simulation-based inference (SBI) approach \citep[e.g.,][]{doi:10.1073/pnas.1912789117}, we infer model parameters using as constraints the CXB, differential AGN number counts in three energy bands, and several observed properties of local AGNs, such as \NH{} distribution, the dependence of the number of obscured and unobscured AGNs on Eddington ratio, and CTK fractions from \textit{Swift}/BAT and NuSTAR. With this approach, we aim to constrain the properties of the dusty tori and the evolution of the local AGN population, while testing the consistency of the radiation-regulated unification model. Finally, we derive the intrinsic CTK fraction, recover the local BHMF and ERDF of obscured and unobscured AGNs separately, and explore the reflection-obscuration and covering factor-Eddington ratio relationships within the synthetic population. Throughout this paper, we use the following cosmological parameters: $H$\textsubscript{0} = 70\,km\,s$^{-1}$\,Mpc$^{-1}$, $\Omega$\textsubscript{m} = 0.3, and $\Omega$\textsubscript{$\Lambda$} = 0.7.

\section{Modelling of AGN population} \label{sec:modelling}
In this section, we present the population synthesis process. We describe the geometrical model we constructed based on the radiation-regulated model, the assumptions made to create the synthetic AGN population, and the tool used to simulate their emission.

\subsection{Black hole mass and Eddington ratio distribution functions} \label{subsec:bhmerfunctions}

To generate the synthetic population, we start from the intrinsic active\footnote{Throughout this work, the BHMF refers to actively accreting (\ER$>10^{-3}$) SMBHs.} black hole mass function (BHMF) and Eddington ratio distribution function (ERDF), which provide the AGN number density in Mpc$^{-3}$ dex$^{-1}$ as a function of redshift and logarithmic mass, $M\mathrm{_{BH}}$, or Eddington ratio, \ER{}, respectively. However, constraints on the evolution of the BHMF and ERDF, particularly for type 2 AGNs, remain relatively limited due to observational and methodological uncertainties (see \citealt{2022ApJS..261....9A} and references therein). Hence, for this work, we use the local BHMF and ERDF from (\citealt{2022ApJS..261....9A}; hereafter A22), who took advantage of the reliably measured $M\mathrm{_{BH}}$ of 678 sources in the 70-month \textit{Swift}/BAT survey, obtained with optical spectroscopy. A22 constrained simultaneously the intrinsic (corrected for survey and measurement biases) BHMF and ERDF in the local Universe ($z < 0.3$) for both unobscured (type 1) and obscured (type 2) AGNs.

The A22 local BHMF, $\Phi_0(\log M\mathrm{_{BH}}) = \frac{\mathrm{d}N}{\mathrm{d}\log M\mathrm{_{BH}}}$ with $N$ being the number of sources, is described by a modified Schechter function:
\begin{equation}\label{eq:BHMF}
\Phi_0(\log M\mathrm{_{BH}}) \propto \ln(10)\, \Phi^* \left(\frac{M\mathrm{_{BH}}}{M_*}\right)^{\alpha +1} \exp \left[-\left(\frac{M\mathrm{_{BH}}}{M_*}\right)^{\beta}\right],
\end{equation}
where $M_*$ is in units of M\textsubscript{\(\odot\)}, $\Phi^*$ is in units of Mpc$^{-3}$ dex$^{-2}$, and $\beta$ changes the slope of the cutoff at high $M_\mathrm{BH}$. The A22 local ERDF, $\xi_0 (\log \lambda_{\mathrm{Edd}}) = \frac{\mathrm{d}N}{\mathrm{d}\log\lambda_{\mathrm{Edd}}}$, is described by a broken power-law:
\begin{equation}\label{eq:ERDF}
\xi_0 (\log \lambda_{\mathrm{Edd}}) \propto \xi^*\left[\left(\frac{\lambda_{\mathrm{Edd}}}{\lambda_*}\right)^{\delta_1}+\left(\frac{\lambda_{\mathrm{Edd}}}{\lambda_*}\right)^{\delta_2}\right]^{-1},
\end{equation}
where $\xi^*$ is in units of Mpc$^{-3}$ dex$^{-2}$ and $\delta_2 > \delta_1$. The integrals of Eq. (\ref{eq:BHMF}) and Eq. (\ref{eq:ERDF}) must provide the same number of sources to ensure consistency between the normalisations of the BHMF and ERDF. Although in principle it would be possible to determine the parameters of the BHMF and ERDF within our framework, for simplicity, we adopt the best-fit values provided in tables E1 and E2 of A22.

To reproduce non-local observables, it is necessary to take into account the evolution of the local BHMF and ERDF. Here, we adopt a simple prescription, where only the number density of active SMBHs evolves with redshift, while the shapes of the BHMF and ERDF remain unchanged (see Sect. \ref{subsec:evolution} for further discussion):
\begin{equation}\label{eq:redshiftfunctions}
\Phi(\log M\mathrm{_{BH}}, z) \propto  \Phi_0(\log M\mathrm{_{BH}})\ e(z),
\end{equation}
with $e(z)$ the evolution factor:
\begin{equation}\label{eq:PDE}
e(z) =  \left\{
         \begin{array}{ l r }
          (1+z)^{p_1},& z<z_c \\
          (1+z_c)^{p_1},& z\geq z_c
         \end{array}
        \right.
\end{equation}
with $z_c$ and $p_1$ free parameters in our model. This prescription implies that each SMBH has a probability of being active that evolves with cosmic time, while SMBH growth is ignored.

In our synthetic population, we include objects with redshifts $z\leq3$, as done in \PaperI{}, because non-blazar AGNs above this redshift do not contribute significantly to the observational constraints (see Sect. \ref{sec:observations}). We sample the BHMF between log($M\mathrm{_{BH}}$/M\textsubscript{\(\odot\)}) = 6.5 and 10.5 and the ERDF between log\ER{} = --3 and 1. Then we calculate the bolometric luminosity, $L_{\rm bol}$, of each object as described in Eq. (5) of A22:
\begin{equation}\label{eq:Lbol}
\log (L\mathrm{_{bol}}/\,\mathrm{erg\,s^{-1}}) = \log \lambda_{\mathrm{Edd}} + \log (M\mathrm{_{BH}}/\mathrm{M_{\odot}}) + 38.18
\end{equation}
We convert $L_{\rm bol}$ into intrinsic 2--10\,keV luminosity:
\begin{equation} \label{eq:Lx}
L\mathrm{_{2-10}} = \frac{L\mathrm{_{bol}}}{\kappa\mathrm{_{2-10}}},
\end{equation}
with $\kappa_{2-10}$ the luminosity-dependent bolometric correction from Table\,1 in \cite{2020A&A...636A..73D}:
\begin{equation}
\kappa_{2-10}(L_\mathrm{bol}) = a\left[1 + \left(\frac{\log (L_{\mathrm{bol}}/L\textsubscript{\(\odot\)})}{b}\right)^{c}\right],
\end{equation}
which was used for tables E1 and E2 in A22.

Randomly drawing objects from the BHMF (or ERDF) can introduce statistical fluctuations at very low redshifts, where the volume is small, and at high masses (or \ER{}), where AGNs are rare. To mitigate these stochastic effects, similarly to what was done in \PaperI{}, we divide the redshift range into redshift bins and draw the same number of objects, $N$\textsubscript{sample}, in each bin. To recover the correct number of objects predicted by the BHMF, $N$\textsubscript{BHMF}, we assign to each AGN a weight of $N$\textsubscript{BHMF}\,/\,$N$\textsubscript{sample}.

\subsection{RefleX} \label{subsec:reflex}

We perform simulations of the AGN model described in Sect. \ref{subsec:model} using the ray-tracing code \textsc{RefleX}\footnote{\url{https://www.astro.unige.ch/reflex/}} 3.0 \citep{2017A&A...607A..31P}, which self-consistently simulates the X-ray emission of AGNs with various geometries, densities, and emission spectra. \textsc{RefleX} performs Monte Carlo simulations to propagate and trace X-ray photons (0.1\,keV -- 1\,MeV) through the circumnuclear material. It allows users to define arbitrary AGN geometries with varying chemical compositions and densities, and to control the geometry and emission spectrum of the X-ray source itself. The code implements key X-ray processes, such as Compton and Rayleigh scattering, photoelectric absorption, and fluorescence. Additionally, \textsc{RefleX} 3.0 \citep{2023ApJ...945...55R} includes dust absorption and scattering based on the Milky Way model from \cite{2003ApJ...598.1026D}. Using \textsc{RefleX}\footnote{The \textsc{RefleX} version we use here includes the additional wedge geometry used for the BLR component (see Sect. \ref{subsec:model}.)}, we generate spectra at selected inclinations, calculate fluxes in specific energy bands, and obtain the line-of-sight \NH{}.

\subsection{AGN model} \label{subsec:model}
\begin{figure}[t]
\centering
\includegraphics[width=\hsize]{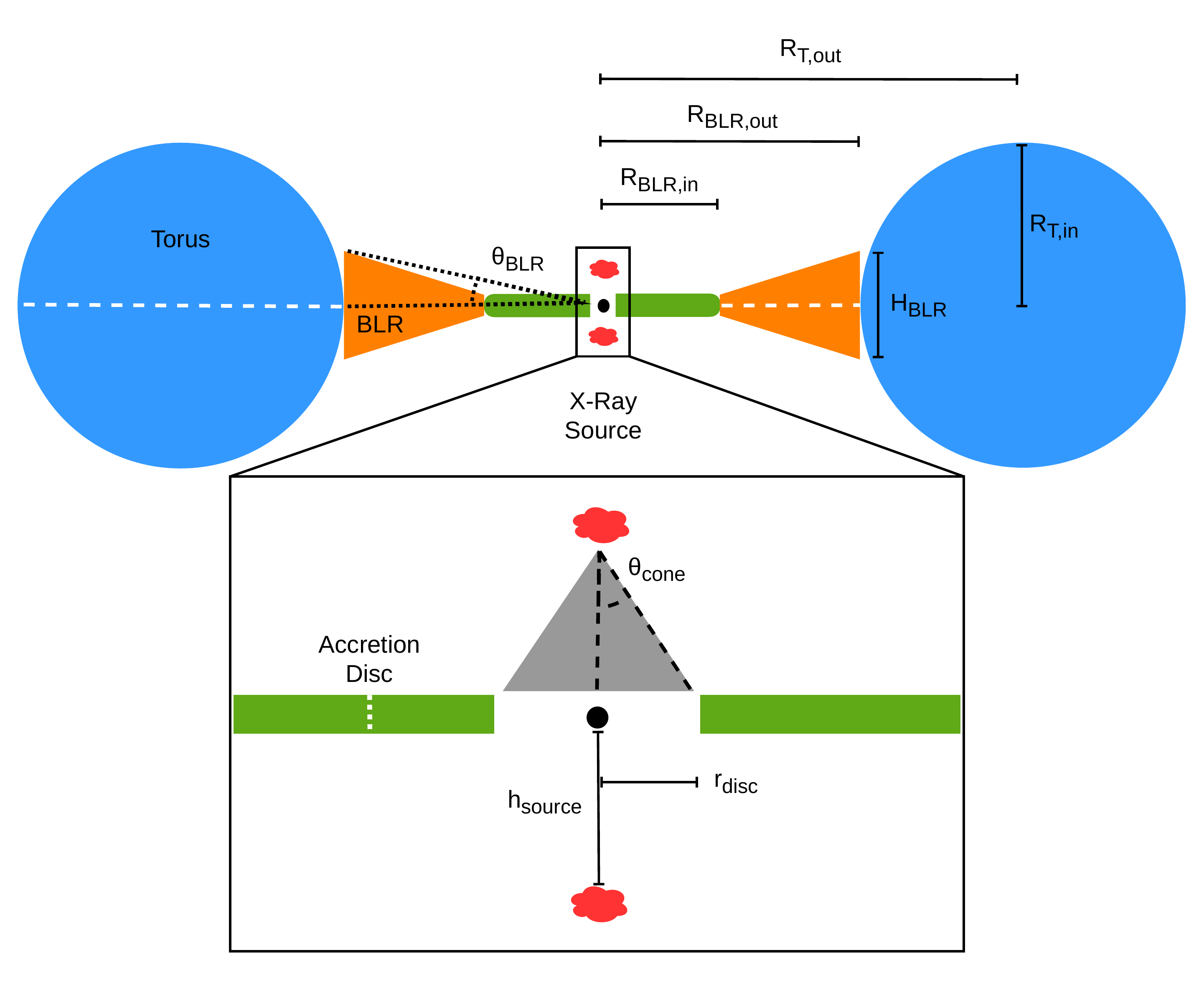}
\caption{AGN geometry used in \textsc{RefleX} (not to scale). The outer radius of the accretion disc is set at the inner radius of the BLR, $R_{\rm disc}$ = $R_{\rm BLR,in}$. The torus centre lies at a distance $R_{\rm T,out}$ from the system centre with a cross-section radius $R_{\rm T,in}$. The inner edge of the torus is set at the outer radius of the BLR so that $R_{\rm T,out}$ -- $R_{\rm T,in}$ = $R_{\rm BLR,out}$. The three white dashed lines indicate the lines across which the column density, \NH\textsubscript{,ref}, is measured for each component. The grey shaded area illustrates the cones defined by the height of the X-ray source, $h_{\rm source}$, and the inner radius of the disc, $r_{\rm disc}$, within which we prevent the sources from emitting since the emission escaping through the centre of the accretion disc would be obstructed by the SMBH. The half-opening angle of the cone, $\theta_{\rm cone}$, is indicated by the black dashed lines. The half-opening angle of the BLR, $\theta_{\rm BLR}$, is indicated by the black dotted lines.
\label{fig:geometry}}
\end{figure}

\begin{table*}
\caption{AGN model components and their properties in \textsc{RefleX}.}\label{tab:model}
\centering
\setlength\extrarowheight{3pt}
\begin{tabular}{lcccccccccc}
\hline\hline
Component & Geometry &
\makecell{$R\rm_{in}$\tablefootmark{a}\\(pc)} &
\makecell{$R\rm_{out}$\tablefootmark{b}\\(pc)} &
\makecell{H\tablefootmark{c}\\(pc)} &
$CF$\tablefootmark{d} &
\makecell{\NH\textsubscript{,ref}\tablefootmark{e}\\(cm$^{-2}$)} &
Dust\,\tablefootmark{f} &
H$_2$\tablefootmark{g} &
\makecell{$L_\mathrm{X}$\tablefootmark{h}\\(erg\,s$^{-1}$)} &
\makecell{Spectrum\\shape\tablefootmark{i}}
\\
\hline
X-ray source & point & - & - & $\pm$10\,$r_{\rm g}$\tablefootmark{j} &-&-&-&-& (\ref{eq:Lx}) & Fig.~\ref{fig:emissionspectrum} \\
Accretion disc & annulus & 6\,$r_{\rm g}$\tablefootmark{j} & (\ref{eq:Rblr}) & -\tablefootmark{k} & - & $10^{29}$ & 0 & 0 & - & - \\
BLR & wedge & (\ref{eq:Rblr}) & (\ref{eq:RsubLbol}) & (\ref{eq:hblr}) & 0.2 & $10^{25}$ & 0 & 0.2 & - & - \\
Torus & torus & (\ref{eq:RsubLbol}),(\ref{eq:CFlambda}) & (\ref{eq:RsubLbol}),(\ref{eq:CFlambda}) & -& (\ref{eq:CFlambda}) & (\ref{eq:Nheq}) & 1 & 0.5 & - & - \\
\hline
\end{tabular}
\tablefoot{This table summarises the properties of an AGN constructed within \textsc{RefleX}. Properties with numerical values are fixed for all AGNs in the synthetic population, while those defined by Eqs. (\ref{eq:Lx})-(\ref{eq:CFlambda}) and Fig. \ref{fig:emissionspectrum} vary across the population. \tablefoottext{a}{Inner radius of each component as defined in Fig. \ref{fig:geometry}.} \tablefoottext{b}{Outer radius of each component as defined in Fig. \ref{fig:geometry}.} \tablefoottext{c}{Height of each component as defined in Fig. \ref{fig:geometry}.} \tablefoottext{d}{Covering factor.} \tablefoottext{e}{Column density of each component along the white dashed lines in Fig. \ref{fig:geometry}.} \tablefoottext{f}{Dust fraction.} \tablefoottext{g}{Molecular hydrogen fraction.}\tablefoottext{h}{Intrinsic X-ray luminosity.} \tablefoottext{i}{X-ray emission spectrum, modelled as a power-law with a high-energy cutoff.} \tablefoottext{j}{$r_{\rm g}$ = G\,$M_{\rm BH}$\,/\,c$^2$ with fixed log($M\mathrm{_{BH}}$/M\textsubscript{\(\odot\)}) = 6.5.} 
\tablefoottext{k}{The annulus is assigned a very small finite height, as required by \textsc{RefleX}; however, this height is negligible compared to the scale of the model.}
}
\end{table*}

Our aim is to construct an AGN model that remains sufficiently simple to parametrise while still being complex enough to incorporate the main AGN components. In Fig. \ref{fig:geometry}, we present the geometrical model that we construct in \textsc{RefleX} (see Sect. \ref{subsec:reflex}), and in Table\,\ref{tab:model}, we summarise the properties of each component. For all simulations, we adopt the matter composition described in \cite{2003ApJ...591.1220L} and assume solar metallicity. A similar model (with the addition of polar obscuration) was built with \textsc{RefleX} and applied on X-ray observations by \cite{2025ApJ...992...64D}. In contrast to \PaperI{}, we directly adopt a complex model without attempting comparisons with simpler models, since we expect even the complex model to be a simplified approximation of reality.

\subsubsection{X-ray source}
AGN X-ray emission most probably originates from a compact corona close to the central black hole \citep[e.g.,][]{2014A&ARv..22...72U}. Here, we use the so-called "lamp-post" geometry \citep[e.g.,][]{1996MNRAS.282L..53M}. Following \PaperI{}, we create a double X-ray point source above and below the centre of the AGN at a height of $h_{\rm source}$ = 10\,$r_{\rm g}$, where $r_{\rm g}$ = G\,$M_{\rm BH}$\,/\,c$^2$ is the gravitational radius. For computational efficiency, we fix $h_{\rm source}$ for all simulated objects at the value corresponding to the lowest SMBH mass in our sample, $M_{\rm BH} =10^{6.5}$ M\textsubscript{\(\odot\)}. This choice results in a slight underestimation of reflection on the accretion disc, especially at certain inclinations and for the rarer objects with high $M_{\rm BH}$. In Appendix \ref{app:massemission}, we show the effect that changing $M_{\rm BH}$ in the geometrical model has on the emission.

As in Paper I, we do not allow the sources to emit  inside the cones defined by $h_{\rm source}$ and the inner radius of the accretion disc, $r$\textsubscript{disc} (as shown in Fig.\ref{fig:geometry}), since the emission that escapes through the centre of the accretion disc would be blocked by the SMBH. The intrinsic luminosity is then corrected by the factor (1 $- \cos\,\theta$\textsubscript{cone}) / 2, where $\theta$\textsubscript{cone} = $\arctan(r$\textsubscript{disc} / $h$\textsubscript{source}) is the half-opening angle of the cones.

For the emission spectrum of the X-ray sources, we choose a power-law with photon index, $\Gamma$, and high-energy cutoff, $E$\textsubscript{C}, dependent on the \ER{} of each AGN. We adopt the $\Gamma$-\ER{} relation found for local changing-state AGNs \citep{2026arXiv260107337J} and the $E$\textsubscript{C}-\ER{} relation found for AGN detected in the 70-month \textit{Swift}/BAT survey \citep{2018MNRAS.480.1819R}. We choose the results of \cite{2026arXiv260107337J} because of the anti-correlation they discovered at low Eddington ratios. We show the empirical relations in Appendix \ref{app:emissionspectrum}.

\subsubsection{Accretion disc}
In addition to the central black hole, the accretion disc is one of the main ingredients of AGNs. Hence, we build a flat, optically thick annulus to model the accretion disc with the geometry described in \PaperI{}. The inner radius of the annulus is fixed at $r$\textsubscript{disc} = 6\,$r$\textsubscript{g}, while the outer radius is set at $R$\textsubscript{disc} = $R$\textsubscript{BLR,in} (see Fig. \ref{fig:geometry}). Similarly to $h$\textsubscript{source}, $r$\textsubscript{disc} is fixed at the value corresponding to $M$\textsubscript{BH} = $10^{6.5}$ M\textsubscript{\(\odot\)}.

\subsubsection{Broad-line region}

The broad-line region (BLR) is a less dense region responsible for the broad emission lines observed from the X-rays to the infrared \citep[e.g.,][]{1988MNRAS.232..539C, 2024ApJ...973L..25X}. To simulate the BLR, we construct a cold, dust-free, flared disc using a wedge geometry. We set the molecular hydrogen fraction to 0.2, and fix the radial column density to 10$^{25}$\,cm$^{-2}$ (see Fig. \ref{fig:geometry}).

It has been well established by reverberation-mapping studies \citep[e.g.,][]{2005ApJ...629...61K, 2013ApJ...767..149B} that the characteristic radius at which the H$\beta$ line is emitted depends on luminosity as $\propto L^{0.5}$, in agreement with simple photoionisation expectations \citep[e.g.,][]{1999ApJ...526..579W}. Accordingly, we let the inner radius of the BLR scale with luminosity:
\begin{equation}\label{eq:Rblr}
\frac{R_\mathrm{BLR,in}}{10\rm{\,lt\,days}} = R_0 \left(\frac{L_\mathrm{2-10}}{10^{43}\,\rm{erg\,s^{-1}}}\right)^{0.5},
\end{equation}
with $R_0$ a free parameter in our model. Here, $L_\mathrm{2-10}$ acts as a proxy for the ionising luminosity. For the outer radius, we adopt the same luminosity dependence as for the inner radius to ensure a self-similar scaling of the BLR structure:
\begin{equation}\label{eq:RsubLbol}
R\mathrm{_{BLR,out}} = R_1 \left(\frac{L_\mathrm{2-10}}{10^{43}\,\rm{erg\,s^{-1}}}\right)^{0.5} \mathrm{pc},
\end{equation}
with $R_1$ a free parameter. The covering factor is fixed at $CF\rm{_{BLR}}=0.2$. This choice is informed by the results of \cite{2017Natur.549..488R}, who find that CTK obscuration remains almost constant at $\sim20\%$ independently of \ER{}, which we attribute to the presence of the BLR. From $R\rm{_{BLR,out}}$ and $CF\rm{_{BLR}}$, we can determine the height of the BLR at $R\rm{_{BLR,out}}$:
\begin{equation}\label{eq:hblr}
    H\rm{_{BLR}} = 2\,R\rm{_{BLR,out}}\,\tan\theta\rm{_{BLR}},
\end{equation}
where the half-opening angle is given by $\theta\rm{_{BLR}}= 90^{\circ} - \arccos$ $CF\rm{_{BLR}}$. 

\subsubsection{Dusty torus}\label{subsubsection:torus}
Finally, we include a cold, dusty, homogeneous torus. We assume a molecular hydrogen fraction of 0.5 \citep[e.g.,][]{2009ApJ...702...63W} and a dust fraction of one, which means that all Fe atoms are in the form of dust grains. We draw the equatorial column density of the tori, \NHeq{}, from a lognormal distribution:
\begin{equation} \label{eq:Nheq}
    \log(N_\mathrm{H,eq}/\rm cm^{-2}) \sim \mathcal{N}(\mu, \sigma^2),
\end{equation}
where $\mu$ and $\sigma$ are the mean and standard deviation, respectively, and free parameters in our model.

The model introduces a dependence of the torus covering factor, $CF$\textsubscript{T}, on \ER{}, in line with the radiation-regulated model \citep[R22,][]{2017Natur.549..488R}. We model $CF$\textsubscript{T} using a sigmoid function:
\begin{equation}\label{eq:CFlambda}
CF_\mathrm{T} \equiv \frac{R_\mathrm{T,in}}{R_\mathrm{T,out}} = CF_\mathrm{BLR} + \left(\frac{0.8}{1 + \exp~[k\, (\log{\lambda_{\mathrm{Edd}}}+1.35)]}\right),
\end{equation}
with $k$ a free positive parameter and the inner and outer radii of the torus, $R_\mathrm{T,in}$ and $R_\mathrm{T,out}$, defined as in Fig. \ref{fig:geometry}. 
The inner edge of the dusty torus, $R_\mathrm{T,out} - R_\mathrm{T,in} = R\rm{_{BLR,out}}$, is set at the outer boundary of the BLR, marking the transition from partially ionised, dust-free gas to the dusty, molecular phase of the circumnuclear material. The radii of the torus, $R_\mathrm{T,in}$ and $R_\mathrm{T,out}$, are then set using Eq. (\ref{eq:RsubLbol}) and (\ref{eq:CFlambda}).

\subsection{Synthetic population} \label{subsec:simulations}
After constructing our AGN model, we use \textsc{RefleX} to perform numerical simulations, generating 24 grids of X-ray spectral models corresponding to all combinations of $\Gamma$ and $E$\textsubscript{C} in Table\,\ref{tab:grids}, with a total of over $6\times10^{6}$ simulated spectra. For computational efficiency, no interpolation is performed along the $\Gamma$ and $E$\textsubscript{C} dimensions. Each grid includes varying ranges for \NHeq{}, $CF_\mathrm{T}$, $R_\mathrm{T,out} - R_\mathrm{T,in}$, $R_\mathrm{BLR,in}$, and $\theta_\mathrm{obs}$, according to Table\,\ref{tab:grids}. The spectrum of any AGN in our population is obtained by interpolating from these grids. For computational efficiency, the grids are computed at $z$ = 0.0001, with redshifting and flux rescaling according to the luminosity distance applied afterwards. In addition, we use \textsc{RefleX} to calculate the line-of-sight column density, \NH{}, at specific $\theta_\mathrm{obs}$. For that, we construct a grid over the same geometrical parameters, neglecting $\Gamma$ and $E_\mathrm{C}$ since these do not affect \NH{}.

\begin{table}
\caption{Dimensions of the \textsc{RefleX} spectral model grids.}\label{tab:grids}
\centering
\setlength\extrarowheight{3pt}
\begin{tabular}{lcc}
\hline\hline
Parameter & Range & Step\\
\hline
$\Gamma$\tablefootmark{a} & 1.6 -- 2.1 & 0.1  \\
$E_C$\tablefootmark{a} (keV) &  150  --  300  & 50  \\
$\log$(\NHeq{}/cm$^{-2}$)\tablefootmark{b} &  20  --  26  &  0.5  \\
$CF_\mathrm{T}$\tablefootmark{c} &  0.2 --  1.0 &  0.1  \\
$\log(R_\mathrm{T,out} - R_\mathrm{T,in}$\,/pc)\tablefootmark{d} & -3.2 --  2.0 &  0.65 \\
$\log (R_\mathrm{BLR,in}$\,/pc)\tablefootmark{d,e} & -4.4 --  1.2 &  0.7  \\
$\theta_\mathrm{obs}$\tablefootmark{f} ($^{\circ}$) &  0  --  90  &  2  \\
\hline
\end{tabular}
\tablefoot{\tablefoottext{a}{Photon index and high-energy cutoff of the emission spectrum.}
\tablefoottext{b}{Equatorial column density of the torus.}
\tablefoottext{c}{Covering factor of the torus.}
\tablefoottext{d}{Outer and inner radius of the BLR.}
\tablefoottext{e}{Only for combinations where $R_\mathrm{BLR,in}< R_\mathrm{T,out} - R_\mathrm{T,in}$.}
\tablefoottext{f}{Orientation of the observer.}
}
\end{table}

We generate a synthetic AGN population by sampling $M_\mathrm{BH}$ and \ER{} from the A22 BHMF and ERDF, a random orientation, $\theta_\mathrm{obs}$, between $0^{\circ}$ and $90^{\circ}$, and an equatorial column density for the torus, \NHeq{}, from a lognormal distribution following Eq. (\ref{eq:Nheq}). To implement the radiation-regulated model, we assign log(\NH{}/\,cm$^{-2}$) = 20 for objects that fall inside the ``forbidden region'' in the \ER{}-\NH{} plane, where AGNs quickly transition from obscured to unobscured due to the high \ER{} blowing away the dusty circumnuclear material \citep{2009MNRAS.394L..89F, 2017Natur.549..488R}.

We use the synthetic population to emulate the surveys from which our observational constraints were derived. For that, we apply the selection functions of each survey to our synthetic population, using the sensitivity curves from the \textit{Swift}/BAT 3-year and 70-month catalogues in \cite{2009ApJ...699..603A} and \cite{2013ApJS..207...19B}, respectively, and from the 40-month NuSTAR serendipitous survey in \cite{2017ApJ...836...99L}. In Appendix \ref{app:selectionfunctions}, we show the sensitivity curves expressed as the sky fraction as a function of 15--55\,keV, 14--195\,keV, and 8--24\,keV flux limit for each study. Using \textsc{RefleX}, we compute these three fluxes for each simulated AGN and obtain the probabilities of detection in each survey. Thus, we ensure that our emulated surveys take into account the selection biases of each survey.

\section{X-ray observational constraints} \label{sec:observations}
To constrain the AGN population, we use various X-ray observations of AGNs, including the CXB spectrum, AGN number counts, and local absorption properties. In Fig. \ref{fig:fits}, we show all datasets described below.

\subsection{Cosmic X-ray background}
As in \PaperI{}, our first observational constraint is the CXB spectrum \citep{1962PhRvL...9..439G}, which comprises the integrated emission of all detected and undetected AGN across cosmic time (see \citealp{2022hxga.book...78B} for a review). The CXB provides a fundamental constraint on the evolution of the AGN population. Successfully reproducing the CXB intensity and shape is essential for our population synthesis model, as the CXB is sensitive to faint and obscured AGNs that may remain undetected in X-ray surveys.

We compare our model to the same compilation of CXB measurements as in \PaperI{} from different instruments: \textit{Chandra}, INTEGRAL (JEM-X, IBIS/ISGRI, SPI), RXTE/PCA, \textit{Swift}/BAT and \textit{Swift}/XRT (\citealp{2017ApJ...837...19C, 2007A&A...467..529C, 2003A&A...411..329R, 2008ApJ...689..666A, 2009A&A...493..501M}). All the data sets (shown in Fig. \ref{fig:fits}a) measure a CXB spectral index of approximately 1.4 below 10\,keV. Residual differences in the normalisation of the CXB among the various measurements are likely driven by differences in spectra cross-calibration, instrumental backgrounds, and cosmic variance \citep[e.g.,][]{2009A&A...493..501M}. We limit our analysis to the 2--100\,keV energy range. That is because at energies above 100\,keV blazars contribute significantly to the CXB spectrum \citep[e.g.][]{2009ApJ...707..778D}, while below 2\,keV normal galaxies contribute increasingly \citep[e.g.,][]{2015MNRAS.451.1892A} and AGN spectra are strongly affected by soft-excess emission and absorption \citep[e.g.][]{2012MNRAS.427..651C}.

\subsection{AGN number counts}
The next set of observational constraints is the differential AGN number counts, defined as the number of sources per flux bin per square degree, in three X-ray energy bands: 14--195\,keV, 8--24\,keV, and 2--10\,keV (Fig. \ref{fig:fits}c-e). The 14--195\,keV number counts are derived in A22 from the 70-month \textit{Swift}/BAT all-sky survey, with a sample of 672 sources at $z \leq 0.3$. A22 fit the counts at fluxes $S_{14-195} \geq 10^{-11.1}\,\mathrm{erg\,s^{-1}\,cm^{-2}}$, obtaining a slope $\mathrm{d} N/ \mathrm{d} S \propto S^{-2.5\pm0.04}$. This is consistent with the results of other hard X-ray surveys, such as those carried out by INTEGRAL and NuSTAR \citep{2010A&A...523A..61K, 2016ApJ...831..185H}.

The 8--24\,keV number counts are taken from a compilation of NuSTAR surveys, with a combined sample of 382 sources detected out to $z=3$ \citep[][hereafter H16]{2016ApJ...831..185H}. In H16, the counts are fitted for fluxes $S_{8-24} \gtrsim 10^{-14}\,\mathrm{erg\,s^{-1}\,cm^{-2}}$, resulting in a slope $\mathrm{d} N/ \mathrm{d} S \propto S^{-2.76\pm0.10}$. H16 show that when the \textit{Swift}/BAT number counts are converted to the same energy band, they underpredict the NuSTAR counts. The discrepancy is attributed to the different redshift ranges probed and thus to the cosmic evolution of AGN number densities.

The 2--10\,keV number counts are based on the ExSeSS catalogue presented in \cite{2023MNRAS.521.1620D} (hereafter D23), which comprises 903 sources detected in \textit{Swift}/XRT observations covering more than 2000\,deg$^2$. D23 fit the counts above $S_{2-10} \geq 10^{-14}\,\mathrm{erg\,s^{-1}\,cm^{-2}}$ and find good agreement with \textit{Chandra} and XMM-\textit{Newton} measurements \citep{2008MNRAS.388.1205G, 2008A&A...492...51M}. They further demonstrate that the \textit{Swift}/BAT and NuSTAR number counts converted into the 2--10\,keV band can be reconciled with the ExSeSS results when appropriate values of obscuration, photon index, and reflection strength are adopted for each survey.

Together, these three energy bands probe different redshift ranges and obscuration regimes. Unlike the 14--195\,keV sample, which primarily probes the local AGN population, reproducing the 8--24\,keV and 2--10\,keV number counts with our model requires contributions from high-redshift sources due to the fainter flux limits. The 2--10\,keV band does not probe the obscured population as efficiently as the harder bands, but provides additional constraints by being more sensitive to absorption.

\subsection{Observed column density distribution}
The third constraint is the observed line-of-sight \NH{} distribution of local AGN, as derived in \cite{2017ApJS..233...17R} (hereafter R17), shown in Fig. \ref{fig:fits}f. This sample includes 731 non-blazar AGN detected by the \textit{Swift}/BAT 70-month survey. With approximately half the AGN classified as obscured, this survey is less biased against high \NH{} sources. Its sensitivity to hard X-rays makes it possible to probe absorption up to the CTK regime, though a selection bias against strongly CTK AGNs remains.

R17 report an \NH{} distribution with a dominant peak in the 20--21 log(\NH{}/cm$^{-2}$) bin and a secondary peak in the 23--24 bin. This suggests that different structures contribute to AGN obscuration: at lower \NH{} the obscuration may be linked to the host galaxy’s plane, while at higher \NH{} it is likely due to the dense circumnuclear material \citep{2008A&A...485..707P}. In our study, we do not model the host galaxy, meaning our simulations lack lowly obscured AGNs. Therefore, as done in \PaperI{}, we compare the number of simulated unobscured objects to all detected AGNs with \NH{} $< 10^{22}$ cm$^{-2}$, assuming a galactic obscuration threshold of log($N$\textsubscript{H,gal}/cm$^{-2}) < 22$ \citep[e.g.,][]{2009MNRAS.400.2050G}.

\subsection{Observed fraction of obscured AGNs as a function of Eddington ratio}
Our fourth constraint concerns the dependence of the fraction of obscured AGNs on Eddington ratio from (\citealt{2022ApJ...938...67R}; hereafter R22). Their sample contains 681 local AGNs detected in the 70-month \textit{Swift}/BAT survey. In order to achieve better completeness, R22 include only the Compton-thin (CTN; commonly $10^{22} \leq$ \NH{} $< 10^{24}$ cm$^{-2}$) AGNs in their obscured AGN distribution. R22 find that the obscured fraction decreases sharply at \ER{} $\geq 10^{-2}$, which agrees with the expected Eddington limit for dusty gas \citep[e.g.,][]{2008MNRAS.385L..43F}. This aligns with the predictions of the radiation-regulated unification model \citep{2017Natur.549..488R}, which suggests that the probability of observing an AGN as obscured depends on the orientation and the accretion rate.

We use the number of obscured and unobscured detected AGNs as a function of \ER{} (shown in Fig. \ref{fig:fits}g-h), instead of their fraction, as they have greater constraining power. The \ER{} in R22 is derived with a constant bolometric correction, $\kappa_{2-10}=20$. For this work, we re-derive the Eddington ratios of the sample using the luminosity-dependent bolometric correction of \cite{2020A&A...636A..73D}. We calculate the uncertainties of the data points from a binomial distribution following \cite{2011PASA...28..128C} as in R22. In Appendix \ref{app:effectkbol}, we compare the effect of the bolometric correction on the obscured fraction and find no significant effect.

\subsection{Observed fraction of Compton-thick AGN}
The final constraint is the observed CTK fraction as a function of the 8--24\,keV limiting flux from three hard X-ray surveys, taken from \cite{2017ApJ...846...20L} (hereafter L17), shown in Fig. \ref{fig:fits}b. In L17, the CTK fraction is derived from eight sources at $z < 0.07$ detected in the 40-month NuSTAR serendipitous survey in the 8--24\,keV energy band. In \cite{2011ApJ...728...58B} (hereafter B11) and \cite{2015ApJ...815L..13R} (hereafter R15), the CTK fraction is derived from 199 and 728 AGNs in the 3-year and 70-month \textit{Swift}/BAT surveys, respectively. L17 converted the \textit{Swift}/BAT limiting fluxes into the 8--24\,keV band assuming a photon index of 1.9, which introduces a systematic uncertainty in the flux limits.

While B11 and R15 find an observed CTK fraction of $<10$\%, L17 find a value more than three times higher at a flux $\sim 30$ times fainter. The observed CTK fractions are a weak constraint compared to the above observational constraints because of the large uncertainties in the measurements and not-modelled incompleteness. Nevertheless, we include them since, along with the CXB, they provide a constraint on the number of CTK objects created by our model.

\section{Results} \label{sec:results}
\subsection{Model parameter determination with SBI} \label{subsec:sbi}

The model presented in Sect. \ref{sec:modelling} contains seven free parameters, while several others are either fixed or constrained from previous studies. The free parameters are summarised in Table\,\ref{tab:priors}, along with their uniform priors, chosen to be uninformative within physically plausible ranges.

We infer the model parameters using simulation-based inference (SBI), as implemented in the Python package sbi \citep{tejero-cantero2020sbi}. Modern SBI methods use deep neural networks to create the joint density of the simulated data and the model parameters. We use the Sequential Neural Posterior Estimation (SNPE) method \citep{2019arXiv190507488G}, which allows direct sampling from the posterior distribution of the model parameters given the observational data. In the first round, the neural density estimator is trained across uniform priors. We then use the covariance matrix of the first posteriors to define a multivariate normal distribution as a proposal for a second round of SNPE. This sequential approach aims at significantly lowering the number of simulations needed for the training of the neural network to converge. We perform 50\,000 simulations in the first round to construct the proposal, and 40\,000 simulations in the second round for retraining.

To ensure that our simulations are comparable to the observational data, we add noise to the simulations. For the CXB spectrum, we added Gaussian noise in each energy bin using the observed variance. For the absorption properties, we applied binomial noise using the model fraction and the number of 'detected' sources per bin. For the differential number counts, we applied Poisson noise using the predicted number of sources in each flux bin.

We use simulation-based calibration (SBC) to evaluate whether the inferred posteriors are well-calibrated. SBC does this by assessing whether the ground-truth parameters fall within posterior confidence intervals at the expected rates. The sbi Python package implements the SBC approach from \cite{2018arXiv180406788T}.

\begin{table}[t]
	\centering
	\caption{Free parameters in the model, their priors, and the median values of their posterior distributions.}
	\label{tab:priors}
	\setlength\extrarowheight{4pt}
	\begin{tabular}{l c c}
		\hline\hline
		Parameter & Prior & Posterior median \\
		\hline
		$\mu$\tablefootmark{1} (log/cm$^{-2}$) & $\mathcal{U}(21.0, 25.0)$ & $23.76\mathrm{_{-0.28}^{+0.28}}$ \\
		$\sigma$\tablefootmark{1} (log/cm$^{-2}$) & $\mathcal{U}(0.1,3.0)$ & $1.6\mathrm{_{-0.15}^{+0.15}}$ \\
		$k$\tablefootmark{2} & $\mathcal{U}(0.1, 4.0)$ & $1.2\mathrm{_{-0.5}^{+1.1}}$ \\
        $R_1$\tablefootmark{3} (pc) & $\mathcal{U}(0.1, 4.0)$ & $1.91\mathrm{_{-0.12}^{+0.14}}$ \\
		$R_0$\tablefootmark{4} (lt\,days) & $\mathcal{U}(0.1, 20.0)$ & $4.71\mathrm{_{-0.18}^{+0.16}}$ \\
		$p_1$\tablefootmark{5} & $\mathcal{U}(0.1, 8.0)$ & $4.12\mathrm{_{-0.05}^{+0.07}}$ \\
        $z_c$\tablefootmark{5} & $\mathcal{U}(0.1, 3.0)$ & $1.00\mathrm{_{-0.03}^{+0.03}}$ \\[2pt]
		\hline
	\end{tabular}
    \tablefoot{The errors represent the 68\% credible interval based on the posterior distributions and $\mathcal{U}(a, b)$ is the uniform distribution between $a$ and $b$. \tablefoottext{1}{Mean, $\mu$, and standard deviation, $\sigma$, of the \NHeq{} distribution in Eq. (\ref{eq:Nheq}).} \tablefoottext{2}{Shape parameter, $k$, in Eq. (\ref{eq:CFlambda}).} \tablefoottext{3}{Normalisation, $R_1$, in Eq. (\ref{eq:CFlambda}).} \tablefoottext{4}{Normalisation, $R_0$, in Eq. (\ref{eq:Rblr}).} \tablefoottext{5}{Slope, $p_1$, and redshift break, $z_c$, of the evolution in Eq. (\ref{eq:PDE}).}}
\end{table}

\subsection{Posterior distributions} \label{subsec:posteriors}

\begin{figure}[t]
\centering
\subfloat{\includegraphics[width = \hsize]{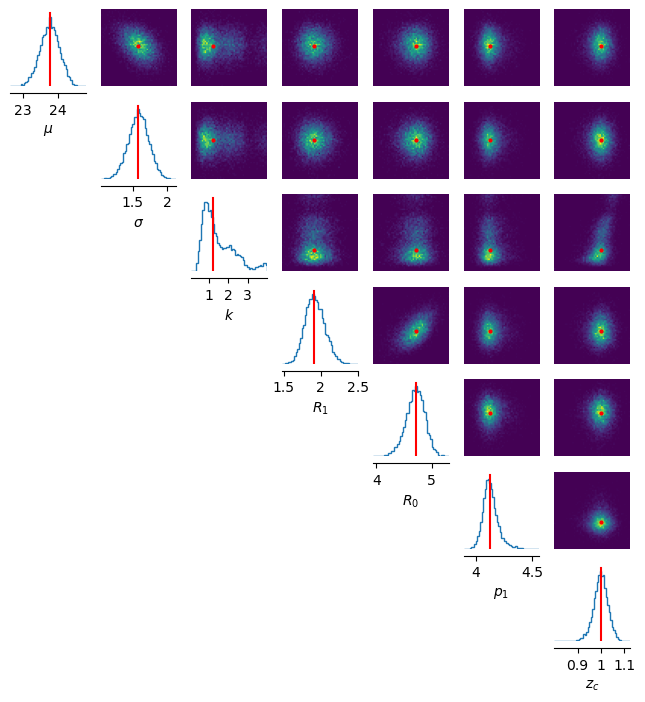}} 
\caption{Posterior distributions of the model parameters produced with SBI. The medians shown in red are reported in Table\,\ref{tab:priors}.}\label{fig:posteriors}
\end{figure}

In Fig. \ref{fig:posteriors}, we present the posterior distributions of the free model parameters, with the corresponding median values reported in Table\,\ref{tab:priors}. The posterior distribution of $\mu$ and $\sigma$ in Eq. (\ref{eq:Nheq}), which essentially set the column densities of the population, appear to be anti-correlated. This is probably due to the need for the model to generate enough high-\NH{} objects. Given the posteriors median values $\mu= 23.76\mathrm{_{-0.28}^{+0.28}}$ and $\sigma = 1.6\mathrm{_{-0.15}^{+0.15}}$, if all dusty tori were observed edge-on, roughly half of them would be CTK.

The posterior distributions of $R_1$ in Eq. (\ref{eq:RsubLbol}) and $R_0$ in Eq. (\ref{eq:Rblr}) appear to be positively correlated, thus if the outer radius of the BLR increases so does the inner one. The posterior of $R_0$ has a median value of $R_0$ = 4.71$\mathrm{_{-0.018}^{+0.016}}$\,lt days, resulting in approximately half the typical BLR sizes inferred from reverberation mapping \citep[e.g.,][]{2005ApJ...629...61K, 2013ApJ...767..149B}. This can be expected since $R_0$ represents the location of the inner boundary of the emitting region rather than the peak of line emission. The parameter $R_1$, which also sets the inner edge of the tori $R_{\rm T,out}-R_{\rm T,in}$, yields a median value of $R_1$ = 1.91$\mathrm{_{-0.12}^{+0.14}}$\,pc. The shape parameter, $k$, which controls the dependence of the torus covering factor on \ER{} in Eq. (\ref{eq:CFlambda}), shows the poorest constraint among all parameters. These results produce an average $R_{\rm T,out}-R_{\rm T,in}$ from $\sim0.5$\,pc at $-3<$ log\ER{} $<-2.5$ to $\sim8$\,pc at $0<$ log\ER{} $<0.5$ and an average $R_\mathrm{T,in}\sim2-3$\,pc across the population. A discussion on these scales is provided in Sect. \ref{subsec:coveringfactor}.

The posterior distributions of the evolution slope and redshift break in Eq. (\ref{eq:PDE}) yield medians $p_1=4.12_{-0.05}^{+0.07}$ and $z_c=1_{-0.03}^{+0.03}$, in broad agreement with the results of XLF studies \citep[e.g.,][see Sect. \ref{subsec:evolution}]{2003ApJ...598..886U, 2010MNRAS.401.2531A, 2014ApJ...786..104U}. Finally, the combination of all the posterior distributions produces a population with $\sim$40\% CTK, $\sim$30\% Compton-thin, and $\sim$30\% unobscured AGNs intrinsically, consistent with recent NuSTAR observations of the local Universe \citep{2025ApJ...978..118B}.

In Appendix \ref{app:coverage}, we show the empirical CDF of the multidimensional SBC ranks, reducing the seven-dimensional parameter space to a scalar using the log probability of the posterior evaluated at each simulated ground-truth parameter set. A well-calibrated inference produces uniformly distributed ranks, so the CDF should follow the 1:1 line. In Fig. \ref{fig:coverage}, the multidimensional CDF is consistent with this expectation, indicating that our posterior is overall well-calibrated. Individual parameter dimensions show good calibration, with the CDFs following the expected uniform distribution.

\subsection{Fit to observational constraints} \label{subsec:fits}

\begin{figure*}[t]
\centering
\subfloat{\includegraphics[width = 0.5\hsize]{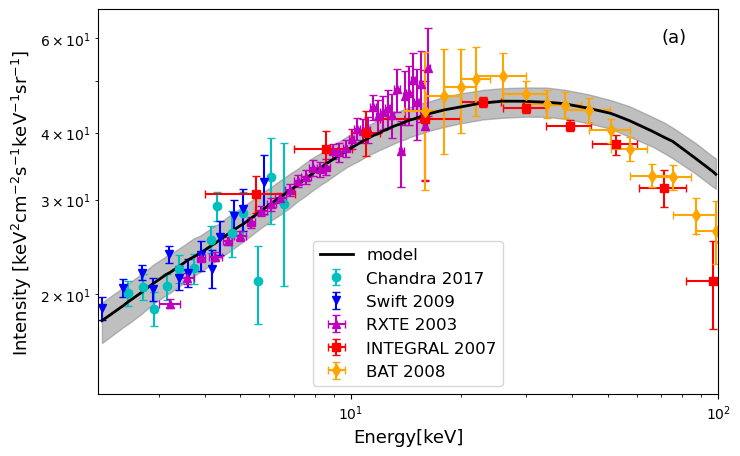}}
\subfloat{\includegraphics[width = 0.37\hsize]{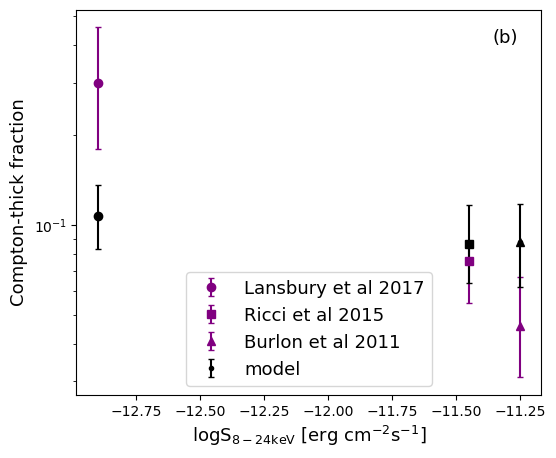}}\\
\subfloat{\includegraphics[width = 0.33\hsize]{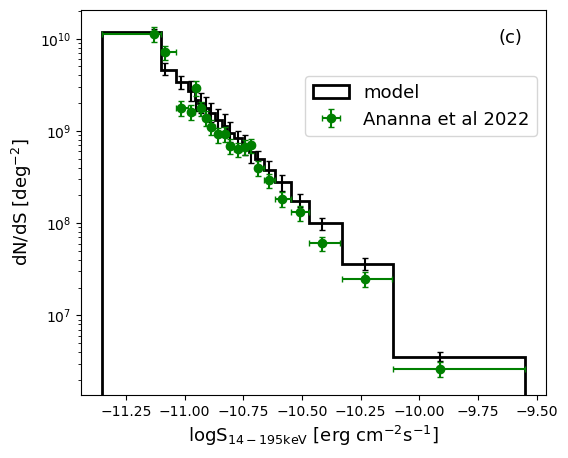}}
\subfloat{\includegraphics[width = 0.325\hsize]{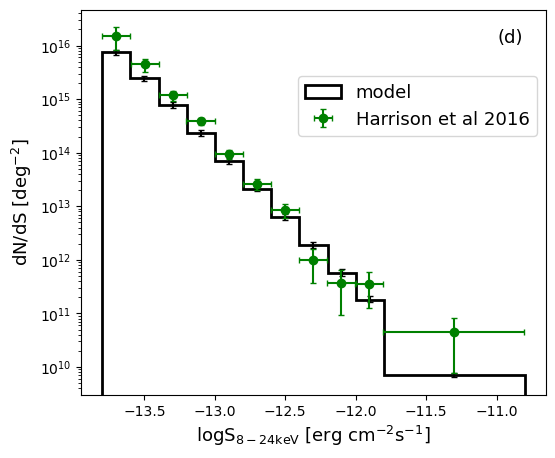}}
\subfloat{\includegraphics[width = 0.325\hsize]{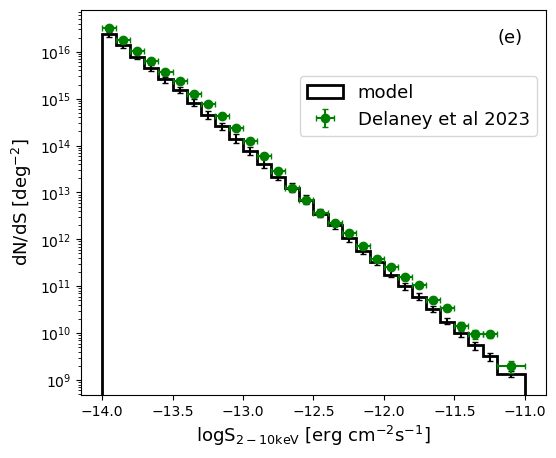}}\\
\subfloat{\includegraphics[width = 0.31\hsize]{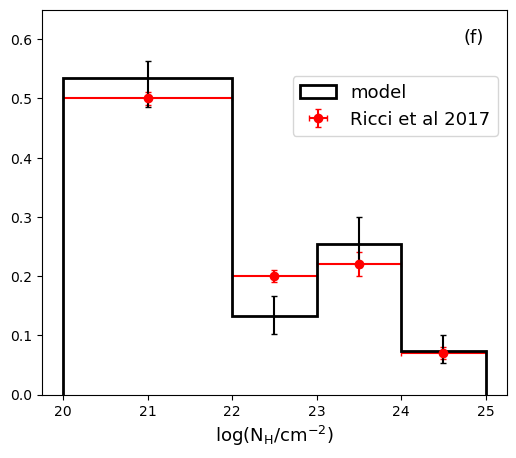}}
\subfloat{\includegraphics[width = 0.328\hsize]{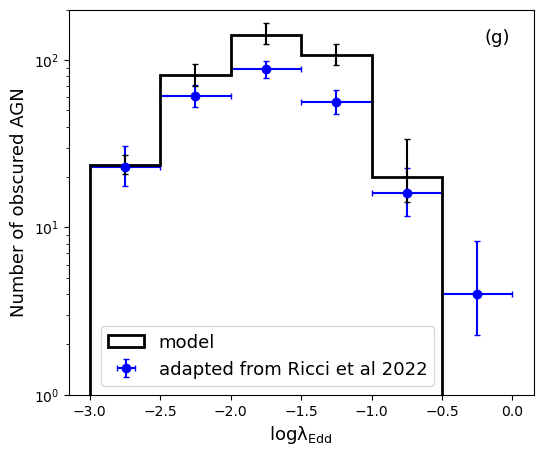}}
\subfloat{\includegraphics[width = 0.328\hsize]{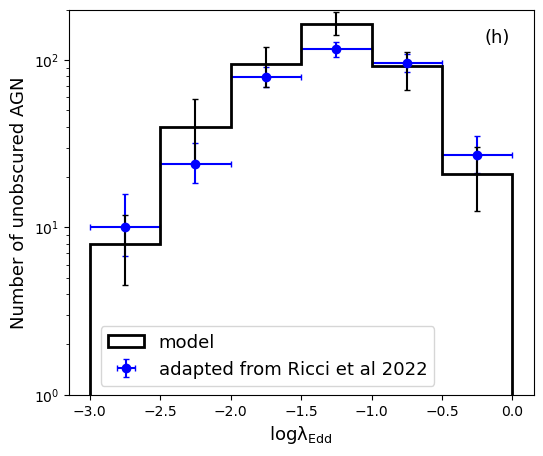}}
\caption{Median model (black) with the grey region and error bars on the model representing the 68\% credible interval based on the posterior distribution of the parameters. (a) CXB spectrum datasets from Chandra \citep[cyan;][]{2017ApJ...837...19C}, INTEGRAL \citep[red;][]{2007A&A...467..529C}, RXTE \citep[magenta;][]{2003A&A...411..329R}, \textit{Swift}/BAT \citep[orange;][]{2008ApJ...689..666A}, and \textit{Swift}/XRT \citep[blue;][]{2009A&A...493..501M}. (b) Observed CTK fractions as a function of 8--24\,keV limiting flux from the: 3-year \textit{Swift}/BAT (triangle; B11), 70-month \textit{Swift}/BAT (square; R15), and 40-month NuSTAR serendipitous survey (circle; L17). Differential number counts from: (c) the 70-month \textit{Swift}/BAT survey in the 14--195\,keV energy band (A22), (d) combined NuSTAR surveys in the 8--24\,keV band (H16), and (e) the ExSeSS catalogue in the 2--10\,keV band (D23). (f) Observed fraction of \NH{} in bins of log(\NH{}) from the 70-month \textit{Swift}/BAT survey (R17). Number of (g) obscured and (h) unobscured AGNs as a function of \ER{} from the 70-month \textit{Swift}/BAT survey (R22), using the bolometric correction from \cite{2020A&A...636A..73D}.} \label{fig:fits}
\end{figure*}

In Fig. \ref{fig:fits}, we show how the synthetic population reproduces the observational constraints given the posterior distributions. Overall, the model reproduces the CXB spectrum, AGN number counts, and absorption properties, although there are some noticeable discrepancies.

The overall intensity of the CXB is very well reproduced across most of the energy range (Fig. \ref{fig:fits}a). However, the model overpredicts the CXB above 60\,keV within 1--2$\sigma$. This discrepancy likely arises from our adopted high-energy cutoff prescription. In particular, the synthetic population is dominated by low-\ER{} objects that, following the $E_\mathrm{C}$-\ER{} relation shown in Fig. \ref{fig:emissionspectrum}, are assigned $E_\mathrm{C} = 300$\,keV, thus causing the predicted high-energy shape. There is a large scatter around the empirical relation found in \cite{2018MNRAS.480.1819R}. Our results indicate that a value of $E_\mathrm{C} \approx 250$\,keV in our prescription may be more appropriate for low-\ER{} objects.

The \textit{Swift}/BAT differential number counts are reproduced mostly within the uncertainties of the data, with the model overpredicting the brightest $S_{14-195}$ flux bins by a factor of $\sim$1.4 (Fig. \ref{fig:fits}c). This behaviour is consistent with several previous population synthesis model predictions (see \cite{2016ApJ...831..185H, 2019ApJ...871..240A} and \PaperI{}), with the causes still under investigation \citep[][]{2023MNRAS.521.1620D}. The model is able to reproduce the number counts from NuSTAR and ExSeSS within 1--2$\sigma$ (Fig. \ref{fig:fits}d-e). However, at the faintest $S_{8-24}$ and $S_{14-195}$ flux bins, the model underpredicts the NuSTAR and ExSeSS counts, by $\sim50$\% and $\sim30$\%, respectively. This shows that the fit makes a compromise in reproducing the faint softer-bands and bright 14--195\,keV counts with a single model. A likely explanation is the limitation we impose in the modelling of the BHMF evolution, suggesting a more physical prescription is required. The inferred evolution parameters of the BHMF appear to be more strongly constrained by the CXB datasets than the number counts datasets, since the CXB normalisation is better reproduced. Such imbalance is not unexpected, as weighting heteroscedastic observables is non-trivial. Moreover, the CXB intensity includes the integrated emission of objects fainter and/or more distant than those contributing to flux-limited number counts. Hence, the CXB is crucial to set a global constraint on the entire underlying population, including its evolution.

The observed CTK fraction is almost constant at a level of $\sim10\%$ irrespectively of the 8--24\,keV limiting flux (Fig. \ref{fig:fits}b). The NuSTAR data point is underpredicted at the 2$\sigma$ level. This systematic discrepancy is shared by previous population synthesis models \citep[e.g.,][]{2014ApJ...786..104U, 2009ApJ...696..110T}, as shown in \cite{2017ApJ...846...20L}. Meanwhile, the observed fraction of \NH{} in bins of log(\NH{}) is well reproduced, capturing both the bimodal form and the relative peak heights mostly within uncertainties (Fig. \ref{fig:fits}f).

The absolute numbers of obscured and unobscured AGNs are overpredicted in the range $-2.5 <$ log\ER{} $< -1$, with a total excess of about $\sim 25\%$ (Fig. \ref{fig:fits}g-h). In the analysis of R22 and its adaptation here, approximately 10\% of sources were excluded from the \textit{Swift}/BAT non-blazar sample due to a lack of $M_\mathrm{BH}$ measurement or complications arising from using broad line $M_\mathrm{BH}$ estimates in obscured sources. This exclusion is not captured by the selection function applied on the synthetic population. Assuming a flat distribution of the excluded sources, the excess drops to $\sim15\%$. Despite the normalisation issue, the dependence of the obscured fraction itself on \ER{} is recovered mostly within uncertainties, as shown in Appendix \ref{app:obscuredfraction}.

\section{Discussion} \label{sec:discussion}
\subsection{Comparison to Paper~I}\label{subsec:comparison}

The AGN population model presented in this work has notable differences compared to the population model presented in \PaperI{}. To start, the two synthetic populations were generated from different intrinsic properties. In \PaperI{}, we sample a $z<3$ XLF whose evolution is determined with the luminosity-dependent density evolution (LDDE) model in \cite{2003ApJ...598..886U}. In contrast, the present population was sampled by a local BHMF and ERDF, whose evolution is inferred self-consistently within this study, although using the simple prescription in Eq. (\ref{eq:PDE}). The inferred evolution in this work is consistent with the evolution found in \cite{2003ApJ...598..886U} using a pure density evolution model (PDE; see Sect. \ref{subsec:evolution} for further discussion).

In both studies, the torus equatorial column density, \NHeq{}, follows a lognormal distribution (Eq. \ref{eq:Nheq}). Here, the \NHeq{} distribution is $\sim$1.6 times broader and shifted 0.2\,dex towards higher values. This occurs despite the additional contribution of BLR obscuration, which was not modelled in \PaperI{}. Additionally, the treatment of the inner edge of the dusty torus is different: in \PaperI{} it is set at the dust sublimation radius with an inferred luminosity dependence of $\propto L_{2-10}^{0.24}$, while in the new model it is set at the outer boundary of the BLR, $R_{\rm BLR,out}$, with a fixed luminosity dependence of $\propto L_{2-10}^{0.5}$. In parallel, \PaperI{} has a fixed torus cross-section radius, $R_{\rm T,in}=1$\,pc, while the new model determines $R_{\rm T,in}$ for each synthetic object from $R_{\rm BLR,out}$ and the \ER{} dependence of the torus covering factor (Eqs. \ref{eq:RsubLbol} and \ref{eq:CFlambda}). Hence, in this work, the combination of all these elements results in more than twice distant tori at luminosities $L_{2-10}>10^{43}$\,erg\,s$^{-1}$ and a double $R_{\rm T,in}$ in average.

Finally, the present model predicts an almost double intrinsic fraction of CTK AGNs (see Sect. \ref{subsec:absorptiondis}), with a corresponding decrease in Compton-thin AGNs, while the intrinsic unobscured population remains the same as in \PaperI{}. These shifts are a direct consequence of our new covering factor–\ER{} relation, which adopts by construction a CTK obscuration of at least 20\% at all accretion levels with the contribution of the BLR.

\subsection{Radiation-regulated unification and properties of the dusty torus} \label{subsec:coveringfactor}

\begin{figure}[t]
\centering
\subfloat{\includegraphics[width=\hsize]{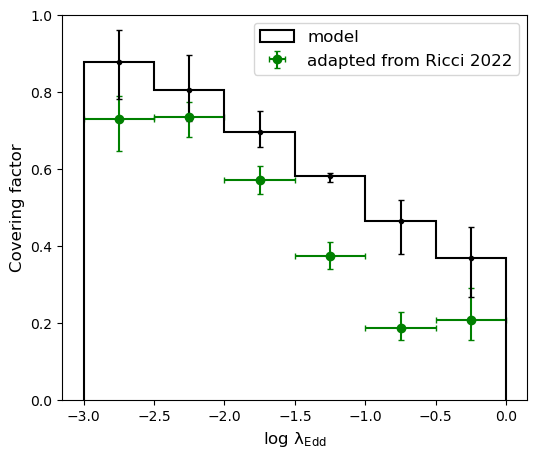}}\\
\caption{Average covering factor as a function of \ER{} predicted by the model (black), with the error bars representing the 68\% credible interval based on the posterior distributions. We compare to the covering factor (CTK plus obscured fractions) adapted from R22 (green), using a luminosity-dependent bolometric correction.
\label{fig:coveringfactor}}
\end{figure}

In this work, we adopt the radiation-regulated unification model proposed by \cite{2017Natur.549..488R}, which links the geometry of the obscuring torus to SMBH accretion. Within this framework, the covering factor of the obscuring material decreases with increasing Eddington ratio, as radiation pressure disperses the dusty circumnuclear material. Fig. \ref{fig:coveringfactor} presents the average torus covering factor as a function of \ER{} predicted by the synthetic population. Our model recovers the trend found in R22 (adapted here using a luminosity-dependent bolometric correction) between -3 $\leq$ log\ER{} $\leq$ 0, with the slope of a linear fit ($a_1 = 0.21\pm0.01$) consistent with that of the data ($a_2 = 0.28\pm0.04$) at the $\sim 1.7\sigma$ level. There is a discrepancy in normalisation mostly alleviated by the fraction of heavily CTK AGNs [log(\NH{}/cm$^{-2}>25$], which are not included in the R22 data, while comprising the majority of CTK objects in our model (see Sect. \ref{subsec:absorptiondis}). It is important to note that while R22 use the CTK and obscured fractions as proxy for the covering factor, our approach infers the covering factor directly from the geometry of the simulated tori.

In our modelling, the inner edge of the dusty torus is defined as $R_\mathrm{T,out}-R_\mathrm{T,in}=R_\mathrm{BLR,out}$, marking the transition from dust-free BLR gas to dusty material. The BLR and torus are treated as contiguous components of the circumnuclear structure, with the radial extent of the BLR scaling with luminosity and the torus covering factor with \ER{}. Hence, the resulting $R_{\rm BLR,out}$ should not be compared to the dust sublimation radius. Instead, here, $R_\mathrm{BLR,out}$ represents the radius at which dust is expected to remain stable within a radiation-regulated environment under the combined effects of gravity and radiation pressure \citep[e.g.,][]{2009MNRAS.394L..89F}. As discussed in \cite{2017Natur.549..488R}, the radiation-regulated unification suggests that most of the obscuring material resides within the gravitational potential of the SMBH, corresponding to typical distances of $1.5-65$\,pc. The inferred inner edge of the tori in our model is well within this scale, with average values of $\sim0.5-8$\,pc for $-3<$ log\ER{} $<0.5$. Combined with the inferred average $R_\mathrm{T,in}\approx2-3$\,pc, the resulting tori lie within the range of mid-infrared interferometric sizes of 0.1--10\,pc, which probe the extent of the warm dust emission \citep[e.g.,][and references therein]{2015ARA&A..53..365N}.

It should be noted that our AGN model has certain key limitations due to the following assumptions. The enforced continuity between the BLR and the torus and their assumed geometrical shapes result in sizes that might not be realistic. Allowing the BLR and torus to have independent radial extents, or introducing an intermediate, possibly multi-phase region, could significantly modify these sizes. A further limitation is the assumption of a fixed BLR covering factor at 0.2, motivated by the constant $\sim$20\% CTK obscuration across \ER{} found in \citet{2017Natur.549..488R}. A more flexible parametrisation, such as allowing the BLR covering factor to vary with \ER{} or luminosity \citep[e.g.,][]{2024Univ...10...29N}, would allow the data to determine whether a constant dust-free component is required. Finally, the assumption that the constant CTK obscuration arises exclusively from the BLR may overlook other contributing physical mechanisms. For instance, a clumpy torus may maintain a high \NH{} close to the equator, even after radiation pressure has cleared the circumnuclear material at larger covering factors \citep[e.g.,][]{2007MNRAS.380.1172H}. At high redshifts, galaxy-scale obscuration could contribute independently of the accretion rate \citep[e.g.,][]{2022A&A...666A..17G}. Polar dust and outflows detected in mid-infrared observations \citep[e.g.,][]{2019MNRAS.489.2177A} may contribute to CTK obscuration in addition to the dusty torus and BLR.

\subsection{Inferred absorption properties of the AGN population} \label{subsec:absorptiondis}

From our AGN population synthesis, we infer an intrinsic CTK fraction of $40\pm3$\%, a significant increase compared to the $21\pm7$\% found in our previous work (\PaperI{}). This higher CTK fraction results from the assumption of a fixed BLR covering factor in our model, which sets a lower limit of 20\% CTK obscuration. Within the CTK population, we find that $64\pm2$\% are heavily CTK [log(\NH{}/\,cm$^{-2}$) > 25], corresponding to $26\pm2$\% of the total AGN population, compared to only $\sim5\%$ in \PaperI{}. This heavy CTK contribution is particularly important as these sources are extremely difficult to detect even in deep hard X-ray surveys. Our result is broadly consistent with local \textit{Swift}/BAT observations corrected for selection biases (B11, R15), which are primarily sensitive to CTK AGNs with log(\NH{}/\,cm$^{-2}$) < 25. Additionally, it is in very good agreement with NuSTAR measurements at $z<0.05$, which report an intrinsic CTK fraction of $35\pm9$\% including heavily CTK sources \citep{2025ApJ...978..118B}. Our findings also align with previous population synthesis predictions at $z \leq 0.1$ \citep{2014ApJ...786..104U, 2015ApJ...802...89B, 2019ApJ...871..240A}.

\begin{figure}[t]
\centering
\subfloat{\includegraphics[width = \hsize]{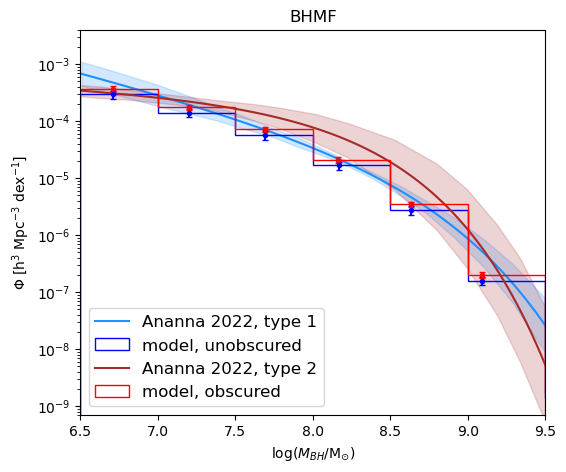}}\\ 
\subfloat{\includegraphics[width = \hsize]{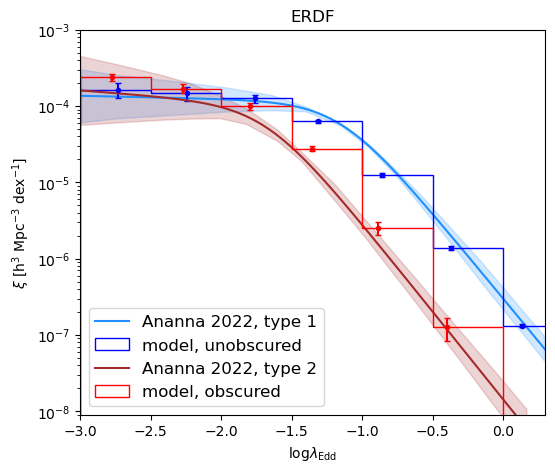}}
\caption{\textit{Top:} BHMF and \textit{bottom:} ERDF of obscured (red points) and unobscured (blue points) objects predicted by the synthetic population compared to the type 1 (light blue line and shaded region) and type 2 (brown line and shaded region) BHMF and ERDF from A22, at $z < 0.3$. The centres of the bins are placed at the median of the corresponding distribution within each bin. The error bars on the model represent the 68\% credible interval based on the posterior distributions.
\label{fig:BHMFtypes}}
\end{figure}

To further test the predictive power of our model, we compare the BHMF and ERDF of obscured and unobscured objects in our synthetic population with the type 1 and type 2 BHMF and ERDF in A22, at $z<0.3$. The synthetic population is generated by the total (absorption-independent) BHMF and ERDF in A22, and the distinction between obscured and unobscured objects emerges after the forward-modelling and parameter inference. Our model recovers the shapes and normalisations of the obscured and unobscured BHMF and ERDF separately, as shown in Fig. \ref{fig:BHMFtypes}. Any deviations, albeit within uncertainties, may be the result of differing definitions between X-ray absorption and optical type in the model and data. The consistency between our results and those from A22 supports our approach and confirms that the differences in the ERDF between obscured and unobscured AGNs can be reproduced by sampling a single parent distribution. The differences then arise from the physical and geometrical properties that determine obscuration within the AGN unification model.

\subsection{Correlation between reflection and obscuration} \label{subsec:reflection}

\begin{figure}[t]
\centering
\includegraphics[width=\hsize]{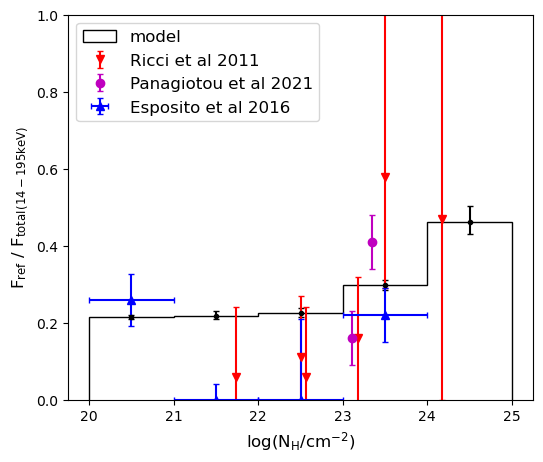}
\caption{Correlation between the average fraction of reflected over total emission, $F$\textsubscript{ref} / $F$\textsubscript{total(14--195keV)}, and the line-of-sight \NH{}, predicted by the model (black). The error bars represent the 68\% credible interval based on the posterior distributions. For comparison, we present the data from INTEGRAL \citep[][red]{2011A&A...532A.102R}, \textit{Swift}/BAT \citep[][blue]{2016A&A...590A..49E}, and NuSTAR \citep[][magenta]{2021A&A...653A.162P} stacked spectra, using the reflection parameter of \textit{pexrav} to estimate $F$\textsubscript{ref} / $F$\textsubscript{total(14--195keV)}. 
\label{fig:reflection}}
\end{figure}

As done in \PaperI{}, in Fig. \ref{fig:reflection}, we use \textsc{RefleX} to qualitatively compare our model predictions to observed trends between the reflection and the line-of-sight \NH{} of stacked hard X-ray spectra. The data from INTEGRAL \citep{2011A&A...532A.102R}, \textit{Swift}/BAT \citep{2016A&A...590A..49E}, and NuSTAR \citep{2021A&A...653A.162P}, show that reflection generally increases with \NH{} for obscured sources. We estimate the fraction of reflected over total emission, $F_{\rm ref}$\,/\,$F_{\rm total}$, in the 14--195\,keV energy band, as proxy for the reflection parameter of the \textit{pexrav} model in XSPEC \citep{1996ASPC..101...17A} reported in the aforementioned studies.

We find that the new model reproduces the trend of increasing reflection with obscuration mostly within the (large) uncertainties. It also predicts the level of reflection in unobscured AGN [log(\NH{}/\,cm$^{-2}$) < 21] in agreement with the observations, whereas the LD+AD model in \PaperI{} predicts an almost double fraction in this range. In the CTK regime, the reflection is reduced to $\sim0.5$, compared to $\sim0.6$ from the previous model. In the intermediate \NH{} range, both models predict similar reflection levels despite their differences. This is not unexpected, as the CXB effectively puts a constraint on the total reflection budget of the synthetic population in both models.

\subsection{Evolution of the fraction of active SMBHs} \label{subsec:evolution}

Constraining the evolution of the active BHMF and ERDF is crucial for understanding SMBH growth across cosmic time. Here, we use a simplistic prescription for the evolution with Eq. (\ref{eq:PDE}) at $z<3$, representing a redshift evolution in the fraction of active SMBHs, while keeping the shapes of the BHMF and ERDF constant. We find a sharp increase in the number density of accreting SMBHs up to $z\approx1$ followed by a plateau. Despite its minimal form, this evolution is able to reproduce the CXB.

The combination of the BHMF and ERDF results in the XLF. Hence, our prescription for the BHMF/ERDF evolution has the same effect as the density evolution term commonly adopted in XLF studies. The posterior of the evolution slope in our model yields a median, $p_1=4.12^{+0.07}_{-0.05}$, broadly consistent with that reported in XLF studies employing pure density and luminosity-dependent density evolution parametrisations \citep{2003ApJ...598..886U, 2010MNRAS.401.2531A, 2014ApJ...786..104U}. Nevertheless, our modelling is not fully physical, and more flexible models, including a mass-dependent evolution, would be required to better describe SMBH growth and accretion-rate distribution at high redshifts \citep[e.g.,][]{2015MNRAS.447.2085S, 2024ApJ...962..152H}.

\section{Conclusions} \label{sec:conclusions}
Previous AGN population synthesis studies have not treated absorption and reflection as physically connected properties. For this work, we build on the self-consistent approach introduced in \PaperI{} and present an updated population synthesis model based on the radiation-regulated unification. For the first time in such studies, we sample an intrinsic local BHMF and ERDF (instead of an XLF) to create a synthetic population, and implement an Eddington-ratio dependent geometry and emission spectrum, when simulating the X-ray emission of AGNs with \textsc{RefleX} (Sect. \ref{sec:modelling}). Our constraints include multiple X-ray observables, such as local absorption properties along with the CXB and AGN differential number counts in three energy bands (Sect. \ref{sec:observations}). We infer the physical properties of the circumnuclear material and the evolution of the active SMBH fraction using SBI (Sect. \ref{subsec:sbi}).

We find that a radiation-regulated-based model is broadly able to explain the wide range of X-ray observations chosen here (Fig. \ref{fig:fits}). The observed absorption properties of local AGNs and the \ER{} dependence of the obscured fraction are well reproduced, although the absolute number of obscured and unobscured detected sources is overpredicted by about 15\%. The model matches the CXB spectrum, except for its shape above $\sim$60\,keV, while the number counts are overpredicted at the bright end in the 14--195\,keV band and underpredicted at the faint end in the 2--10 and 8--24\,keV bands. The average inferred inner edge of the torus spans $\sim0.5-8$\,pc between $-3<$ log\ER{} $<0.5$, which agrees with the scales expected in a radiation-regulated scenario \citep{2017Natur.549..488R} and measured sizes from mid-infrared observations \citep[e.g.,][and references therein]{2015ARA&A..53..365N}. The model predicts an intrinsic CTK fraction of $40\pm3$\%, with $64\pm2$\% of them having log(\NH{}/cm$^{-2}$)$>25$. This CTK fraction is almost twice as high as the one predicted in \PaperI{}, mostly due to the addition of obscuration from the BLR. However, the new prediction is consistent with recent observations in the local Universe \citep{2025ApJ...978..118B} and with several previous population synthesis models at $z \leq 0.1$ \citep{2014ApJ...786..104U, 2015ApJ...802...89B, 2019ApJ...871..240A}. 

Our model recovers the observed anti-correlation between the torus covering factor and the Eddington ratio from the torus geometry itself (Fig. \ref{fig:coveringfactor}) and the local BHMFs and ERDFs of obscured and unobscured AGNs (Fig. \ref{fig:BHMFtypes}), while it is in general agreement with the observed correlation between reflection and obscuration (Fig. \ref{fig:reflection}). Nevertheless, this work makes several assumptions about the obscuring material, including the geometry of the BLR and dusty torus, their homogeneous distribution, and the lack of other potential obscurers, which can be improved in future studies. Despite using a simplistic prescription for the BHMF evolution by controlling only the active fraction of SMBHs, we are able to reproduce the intensity of the CXB from our locally-sampled population synthesis. This demonstrates the future potential for combining the local BHMF and ERDF, which are available for both type 1 and type 2 AGN, with more complex evolutionary models to attempt to constrain SMBH growth history within our framework.
          
\bibliographystyle{aa}
\bibliography{Bib}

\begin{appendix}

\section{Effect of SMBH mass on emission} \label{app:massemission}

\begin{figure}[h!]
\centering
\subfloat{\includegraphics[width = \hsize]{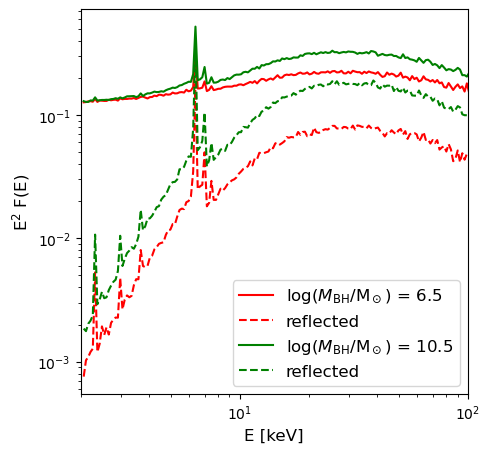}}\\ 
\subfloat{\includegraphics[width = \hsize]{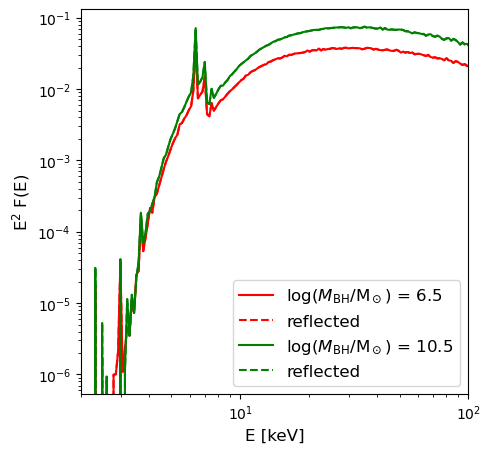}}
\caption{Total (solid lines) and reflected (dashed lines) simulated emission by \textsc{RefleX} for two objects with the same properties and log(\NHeq{}/cm$^{-2})=24$, but for log($M_\mathrm{BH}$/M\textsubscript{\(\odot\)}) = 6.5 (red lines) and log($M_\mathrm{BH}$/M\textsubscript{\(\odot\)}) = 10.5 (green lines). \textit{Top:} face-on. \textit{Bottom:} edge-on, where the spectrum becomes reflection-dominated and the solid and dashed lines are identical. \label{fig:massemission}}
\end{figure}

The height of the X-ray corona, $h_\mathrm{source} = 10\ r_\mathrm{g}$, and the inner radius of the accretion disc, $r_\mathrm{disc} = 6\ r_\mathrm{g}$, are proportional to the SMBH mass, $M_\mathrm{BH}$. For this work, we fix $h_\mathrm{source}$ and $r_\mathrm{disc}$ for log($M_\mathrm{BH}$/M\textsubscript{\(\odot\)}) = 6.5. In Fig. \ref{fig:massemission}, we show the effect on emission when changing the mass to the maximum one in the population, log($M_\mathrm{BH}$/M\textsubscript{\(\odot\)}) = 10.5.

\section{Emission spectrum relations} \label{app:emissionspectrum}

\begin{figure}[h!]
\centering
\subfloat{\includegraphics[width = \hsize]{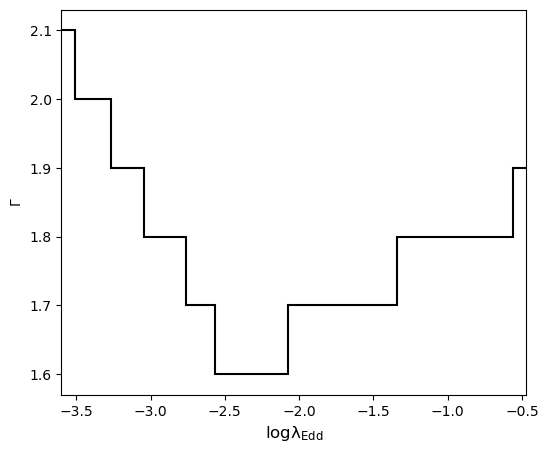}}\\ 
\subfloat{\includegraphics[width = \hsize]{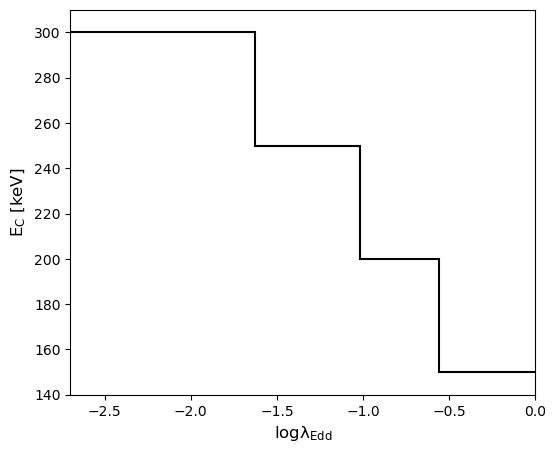}}
\caption{\textit{Top:} Empirical relation between the photon index and the Eddington ratio of local changing-state AGNs from \cite{2026arXiv260107337J}, used here to define a step function for $\Gamma$. \textit{Bottom:} Empirical relation between the high-energy cutoff and the Eddington ratio of sources detected in the 70-month \textit{Swift}/BAT survey from \cite{2018MNRAS.480.1819R}, used here to define a step function for $E_\mathrm{C}$.
\label{fig:emissionspectrum}}
\end{figure}

\clearpage

\section{Selection functions} \label{app:selectionfunctions}

\begin{figure}[h!]
\centering
\includegraphics[width=\hsize]{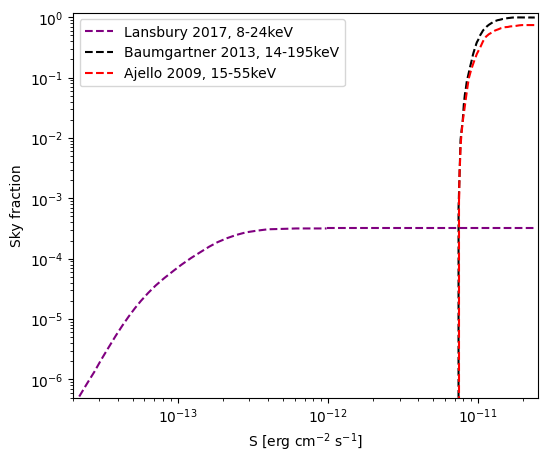}
\caption{Sky fraction as a function of 15--55\,keV, 14--195\,keV and 8--24\,keV flux limits representing the sensitivity curves of the 3-year (red line; from \citealp{2009ApJ...699..603A}) and 70-month (black line; from \citealp{2013ApJS..207...19B}) \textit{Swift}/BAT surveys used in B11 and R15, respectively, and the 40-month NuSTAR serendipitous survey (purple line; from \citealp{2017ApJ...836...99L}) used in L17.
\label{fig:sensitivity}}
\end{figure}

\section{Effect of bolometric correction on obscured fraction} \label{app:effectkbol}

\begin{figure}[h!]
\centering
\subfloat{\includegraphics[width = \hsize]{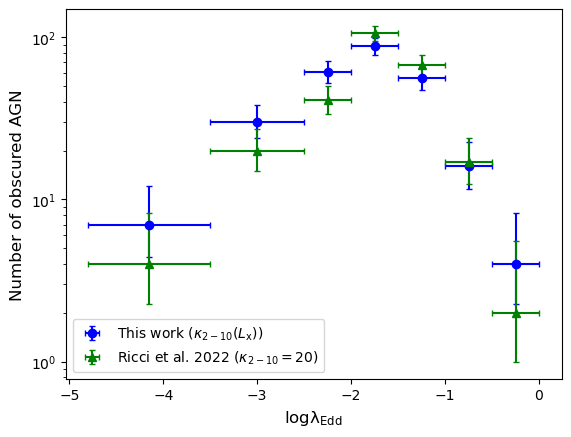}} \\
\subfloat{\includegraphics[width = \hsize]{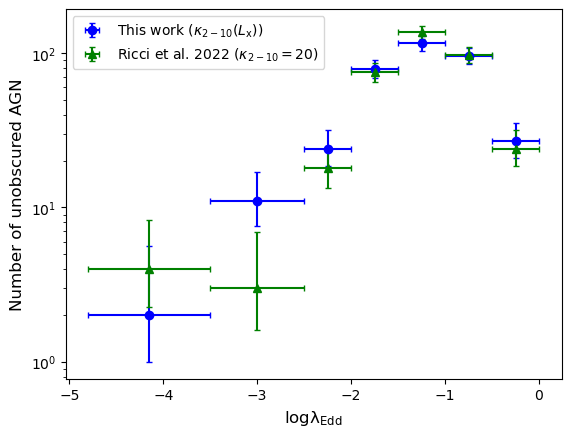}} \\
\subfloat{\includegraphics[width = \hsize]{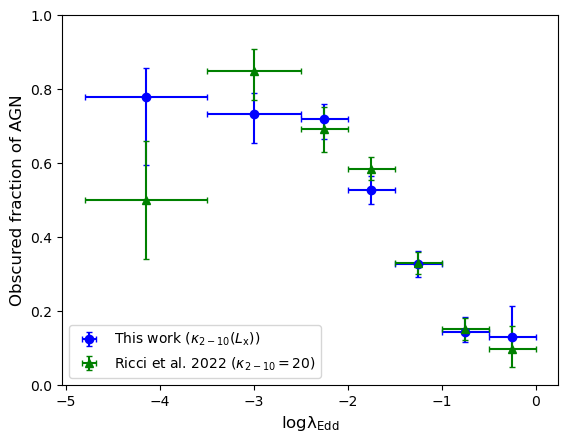}} 
\caption{\textit{Top:} number of obscured AGN as a function of log\ER{} in the 70-month \textit{Swift}/BAT survey with a constant bolometric correction (green points; \citealt{2022ApJ...938...67R}) versus the luminosity-dependent correction from \cite{2020A&A...636A..73D} used in this work (blue points). \textit{Middle:} number of unobscured AGN as a function of log\ER{}. \textit{Bottom:} fraction of obscured AGN as a function of log\ER{}.}\label{fig:obsvskbol}
\end{figure}
\clearpage

\section{Simulation-based calibration} \label{app:coverage}

\begin{figure}[h!]
\centering
\subfloat{\includegraphics[width = \hsize]{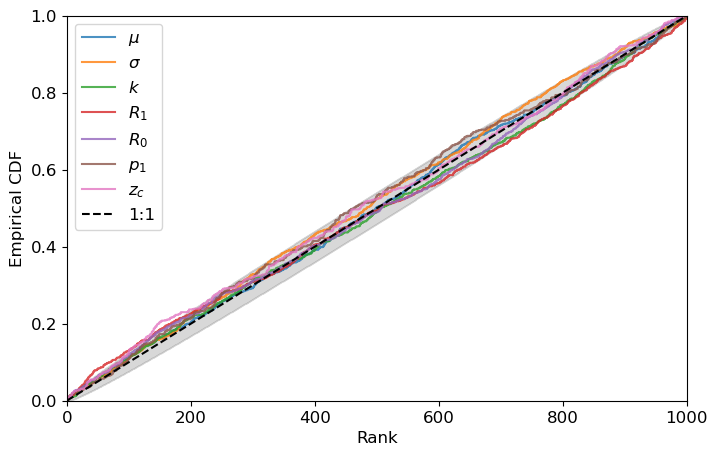}}\\
\subfloat{\includegraphics[width = \hsize]{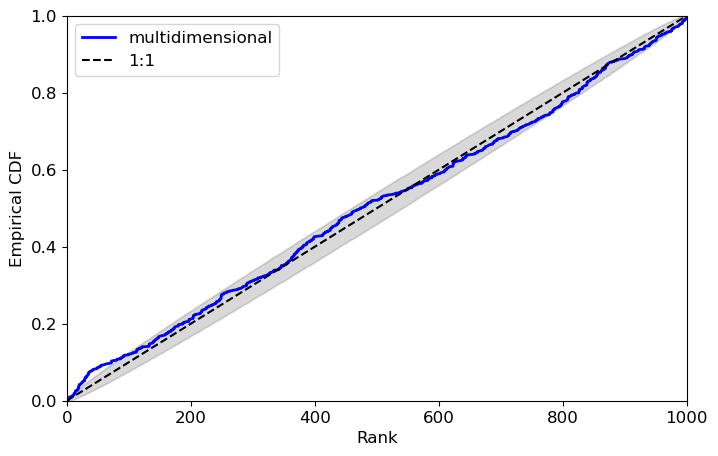}}
\caption{Empirical CDFs of the ranks from \textit{top:} SBC of each dimension, $\mu$ (light blue), $\sigma$ (orange), $k$ (green), $R_1$ (red), $R_0$ (mauve), $p_1$ (brown), and $z_\mathrm{c}$ (pink), and \textit{bottom:} multidimensional SBC (blue). The expected uniform distribution is represented by the black dashed line and its 95\% credible interval by the grey region.}\label{fig:coverage}
\end{figure}

\section{Fraction of obscured AGNs as a function of Eddington ratio} \label{app:obscuredfraction}

\begin{figure}[h!]
\centering
\subfloat{\includegraphics[width = \hsize]{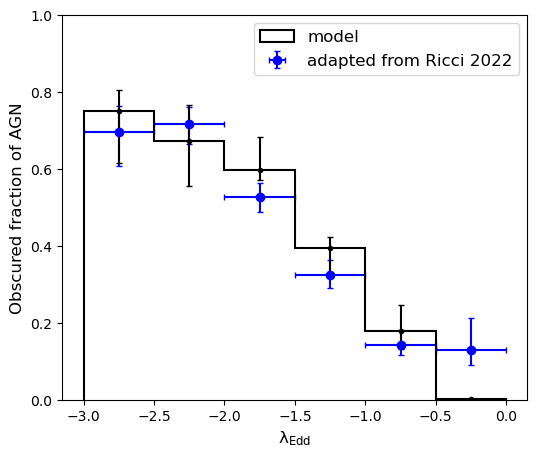}}
\caption{Model prediction (black) for the fraction of obscured AGNs as a function of Eddington ratio. The error bars on the model represent the 68\% credible interval based on the posterior distributions. For comparison, the data are adapted from R22, using the bolometric correction from \cite{2020A&A...636A..73D} (blue).}\label{fig:obscuredfraction}
\end{figure}

\end{appendix}

\end{document}